\documentclass[amsmath,amssymb,prb,twocolumn,superscriptaddress,longbibliography]{revtex4-2}

\usepackage[final]{pdfpages}
\usepackage{graphicx}
\usepackage[utf8]{inputenc}
\usepackage{color}
\usepackage{amsmath,amsfonts,amssymb}
\usepackage{lpic}
\usepackage{ctable}
\usepackage{tabularx}
\usepackage{hyperref}

\newcommand{\dipc}{Donostia International Physics Center (DIPC), E-20018, Donostia-San Sebasti\'an, Spain}
\newcommand{\ikerbasque}{IKERBASQUE, Basque Foundation for Science, E-48013, Bilbao, Spain}
\newcommand{\cfm}{Centro de F\'{\i}sica de Materiales (CFM) CSIC-UPV/EHU, E-20018, Donostia-San~Sebasti\'an, Spain}

\newcommand{\Figref}[1]{Fig.~\ref{#1}}
\newcommand{\Eqref}[1]{Eq.~(\ref{#1})}
\newcommand{\Secref}[1]{Sec.~\ref{#1}}
\newcommand{\ie}{\textit{i.e.}}

\makeatletter
\patchcmd{\@outputpage@head}{\@ifx{\LS@rot\@undefined}{}{\LS@rot}}{}{}{}
\makeatother

\begin{document}

\title{Electron beam splitting effect with crossed zigzag graphene nanoribbons\\ in high-spin metallic states}

\author{Sofia Sanz}
\email{sofia.sanzwuhl@ehu.eus}
\affiliation{\dipc}
\affiliation{\cfm}
\author{G\'eza Giedke}
\affiliation{\dipc}
\affiliation{\ikerbasque}
\author{Daniel S\'anchez-Portal}
\affiliation{\cfm}
\author{Thomas Frederiksen}
\email{thomas\_frederiksen@ehu.eus}
\affiliation{\dipc}
\affiliation{\ikerbasque}

\date{\today}

\begin{abstract}
Here we analyze the electron transport properties of a device formed of two crossed graphene nanoribbons with zigzag edges (ZGNRs) in a spin state with total magnetization different from zero. While the ground state of ZGNRs has been shown to display antiferromagnetic ordering between the electrons at the edges, for wide ZGNRs--where the localized spin states at the edges are decoupled and the exchange interaction is close to zero--, in presence of relatively small magnetic fields, the ferromagnetic (FM) spin configuration can in fact become the state of lowest energy due to the Zeeman effect. In these terms, by comparing the total energy of a periodic ZGNR as a function of the magnetization per unit cell we obtain the FM-like solution of lowest energy for the perfect ribbon, the corresponding FM-like configuration of lowest energy for the four-terminal device formed of crossed ZGNRs, and the critical magnetic field needed to excite the system to this spin configuration. By performing transport calculations, we analyze the role of the distance between layers and the crossing angle of this device in the electrical conductance, at small gate voltages. The problem is approached employing the mean-field Hubbard Hamiltonian in combination with non-equilibrium Green's functions.
We find that ZGNR devices subject to transverse magnetic fields may acquire a high-spin configuration that ensures a metallic response and tunable beam splitting properties,  making this setting promising for studying electron quantum optics with single-electron excitations.
\end{abstract}

\maketitle

\section{Introduction}

The increasing interest in graphene nanoribbons (GNRs) for molecular-scale electronic and spintronic devices has emerged because it is well known that they inherit some of the exceptional properties of graphene while having tunable electronic properties, such as the dependence of the band gap on their width and edge topology \cite{Son2006a}, and the appearance of $\pi$-magnetism \cite{ OtDiFr.22.Carbonbasednanostructures}, absent in pure two-dimensional (2D) graphene.
Moreover, these systems have been shown to be a remarkable platform for electron quantum optics, where the electrons propagating coherently in these ballistic conductors resemble photons propagating in optical waveguides.
On the one hand, it has been shown that electrons can propagate without scattering over large distances of the order of $\sim$100 nm in GNRs \cite{Minke2012, Baringhaus2014, Aprojanz2018}. On the other hand, ballistic transport in ZGNRs can be fairly insusceptible to edge defects as a consequence of the prevailing Dirac-like behavior, that make the electronic current flow maximally through the central region of the ribbon \cite{Zarbo2007}.
Furthermore, with the advent of bottom-up fabrication techniques, long samples of GNRs free of defects can now be chemically realized via on-surface synthesis, as demonstrated in the seminal works by Cai \emph{et al.} for armchair GNRs \cite{Cai2010} and by Ruffieux \emph{et al.} for ZGNRs \cite{Ruffieux2016}.

It is known that the ground state of ZGNRs corresponds to a ferromagnetic (FM) ordering of spins along the edges and antiferromagnetic (AFM) ordering between the edges \cite{Fujita1996, Lee2005}, that is, with total spin projection per unit cell equal to zero, $S_z=0$.
In this configuration, the magnetic instabilities of the states localized at the edges coming from the flat bands of ZGNRs open a band gap due to Coulomb repulsion in the otherwise metallic ribbons \cite{Son2006b}.
The opening of the band gap and the edge states associated with the AFM coupling in ZGNRs have been confirmed by experiments, where the magnetic order has been shown to be stable up to room temperature \cite{Magda2014, Li2014a}. The spin-polarized states localized at the edges are coupled such that there is an energy penalty to excite the AFM ground state to the FM state (exchange interaction).
In the case of wider ZGNRs, the AFM and FM solutions are close in energy (small exchange interaction) due to the decoupling of the localized edge states, as they decay exponentially towards the center of the ribbon \cite{Fujita1996, Nakada1996, Wakabayashi1999, Miyamoto1999,PhysRevLett.102.227205}. 
In this case, the FM solution can be favored due to the Zeeman energy under a relatively small magnetic field. The presence of a net spin-polarization, in absence of transition metals or heavy atoms, makes these structures privileged for spintronics due to the weak spin scattering in pure carbon-based systems \cite{Rocha2005, OtDiFr.22.Carbonbasednanostructures}.
For instance, the intrinsically weak spin-orbit and hyperfine couplings in graphene lead to long spin coherence and relaxation times \cite{Han2011} as well as a large spin-diffusion length that is expected to reach $\sim 10 \mu$m even at room temperature \cite{Tombros2007}.

Recently, devices formed of crossed GNRs have been predicted to behave as perfect beam splitters, where the injected electron beam is divided into two out of the four arms with near 50-50 probability and zero backscattering \cite{Lima2016, Brandimarte2017, Sanz2020}. Furthermore, the particular case of devices formed of crossed ZGNRs is even more interesting, since they can create a spin-polarizing scattering potential \cite{Sanz2022} where the device can work as a spin-polarizing beam splitter.  Following these ideas for electron quantum optics applications, a Mach--Zehnder-like interferometer in a GNRs-network has recently been proposed \cite{Sanz2023}.
In terms of their feasibility, manipulation of GNRs in STMs \cite{Koch2012, Kawai2016} has opened the possibility to build 2D multi-terminal GNR-based electronic circuits \cite{Jiao2010}. The spin properties of such devices can be addressed by measuring with spin-polarized STMs \cite{Wortmann2001, Brede2023} and probed by shot-noise measurements \cite{Burtzlaff2015}. For instance, a device formed of two crossed ZGNRs has been experimentally realized with the control over the crossing angle reaching a precision of 5$^\circ$ \cite{Wang2023}.

While in previous works only the AFM regime has been explored, other spin configurations can appear and show interesting spin-polarized transport properties.
For instance, in contrast to the AFM case, the FM band structure of periodic ZGNRs does not show a band gap around the Fermi level, which makes this regime interesting since there is conduction of electrons at the Fermi level.
Given the metallic character of the FM-like spin configuration, one can envision to generate a minimal excitation in the device with only one particle and no hole (a leviton) \cite{Levitov1996, Ivanov1997, Keeling2006, Dubois2013} by applying a small voltage pulse \cite{Assouline2023}.

Here we analyze the functioning of an electronic beam splitter built with two crossed ZGNRs (of width 30 carbon atoms across) in a FM-like configuration, \ie, where the total magnetization of the device is different from zero. To describe the spin physics of the system we employ the Hubbard Hamiltonian in the mean-field approximation (MFH) \cite{Hubbard1963}. The main complexity of the modeling lies in the description of the coupling between ZGNRs at the crossing, for which we use a Slater-Koster parametrization \cite{Slater1954} that has shown to be in good agreement with other more accurate descriptions, such as density functional theory \cite{Sanz2020}. By employing this simple, yet powerful description based on single-electron physics we are able to explore large systems composed by $\sim 8000$ atoms.

The manuscript is structured as follows: In \Secref{sec:methods} we explain in detail the theoretical methods employed in this work (MFH Hamiltonian and NEGF formalism), in \Secref{sec:results} we present the obtained results for a device formed of two crossed wide ZGNRs in its FM-like configuration, and finally, the conclusions are provided in \Secref{sec:conclusions}.

\section{Methods}
\label{sec:methods}

\begin{figure}
    \centering
    \includegraphics[width=\columnwidth]{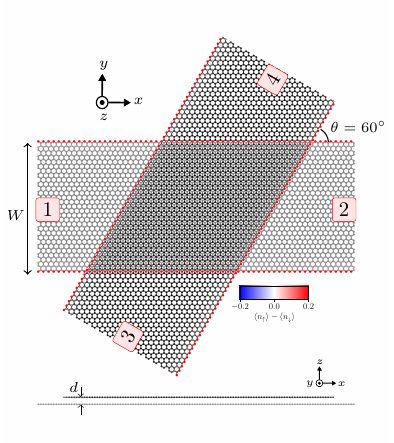}
    \caption{Top and side view of the device geometry with spin density distribution. The size of the blobs is proportional to the magnitude of the spin polarization, $\langle n_{\uparrow}\rangle-\langle n_{\downarrow}\rangle$, and the color depicts the sign of the spin polarization as indicated by the colorbar placed as an inset figure. The four numbered electrodes are indicated in red squares. The crossing angle between the ribbons in this geometry is $\theta=60^{\circ}$. The layers are separated by a distance $d$. The width ($W$) of the ribbons is 30 atoms across.}
    \label{fig:device}
\end{figure}

The system of study is composed by two infinite crossed ZGNRs placed one on top of the other separated by an inter-ribbon distance $d$, with a relative crossing angle of around $\theta=60^\circ$, as shown in \Figref{fig:device}. Here, the semi-infinite electrodes are indicated by red squares numbered 1 to 4.

To describe the $\pi$-electrons, responsible for the spin polarization and the transport phenomena in the system in presence of Coulomb repulsion, we employ the MFH Hamiltonian \cite{Hubbard1963} with a single $p_z$ orbital per site, 
\begin{equation}\label{eq:Hubbard-Ham}
H_\mathrm{MFH} = \sum_{ij,\sigma}t_{ij}c^{\dagger}_{i\sigma}c_{j\sigma}^{\phantom{\dagger}} + U\sum_{i,\sigma}n_{i\sigma}\left\langle n_{i\overline{\sigma}}\right\rangle.
\end{equation}
Here $c_{i\sigma}$ ($c^\dagger_{i\sigma}$) is the annihilation (creation) operator of an electron at site $i$ with spin $\sigma=\{\uparrow,\downarrow\}$ and $n_{i\sigma}=c_{i\sigma}^\dagger c_{i\sigma}^{\phantom{\dagger}}$ the corresponding number operator.
The tight-binding parameters $t_{ij}$ are described by Slater--Koster two-centre $\sigma$- and $\pi$-type integrals between two $p_{z}$ atomic orbitals \cite{Slater1954} as used previously for twisted-bilayer graphene \cite{ TramblydeLaissardiere2010} and crossed GNRs \cite{Sanz2020, Sanz2022, Sanz2023}.
$U$ accounts for the Coulomb interaction between two electrons occupying the same $p_z$ orbital.
The total Hamiltonian $H_T$ is the  composition of the device Hamiltonian $H_D$, the electrodes Hamiltonian for the periodic leads $H_{\alpha}$, and the coupling between these two $H_{\alpha D}$, \ie, $H_T=H_D + \sum_{\alpha}\left(H_{\alpha} + H_{\alpha D}\right)$. More details for the implementation can be found in Refs. \cite{Sanz2022, Sanz2023, dipc_hubbard}.

As the junction between the ribbons breaks the translational invariance of the perfect ZGNRs, we use the Green's function \cite{Keldysh1965, Kadanoff1962} formalism to solve the Schr\"odinger equation for the open quantum system.
Details of the implemented MFH model with open boundary conditions \cite{dipc_hubbard} can be found in the supplemental material of Ref.~\cite{Sanz2022}.

The transport properties are analyzed by computing the transmission probabilities per spin index $\sigma=\lbrace\uparrow,\downarrow \rbrace$, between the different pairs of terminals as a function of the electron energy $E$ from the Landauer-B\"uttiker formula \cite{Buettiker1985, Buettiker1990},
\begin{equation}\label{eq:transmission}
    T^{\sigma}_{\alpha\beta}(E) = \mathrm{Tr}\left[\mathbf{\Gamma}^{\sigma}_{\alpha}\textbf{G}^{\sigma}\mathbf{\Gamma}^{\sigma}_{\beta}\textbf{G}^{\sigma\dagger} \right],
\end{equation}
where $\textbf{G}$ is the retarded Green's function and $\mathbf{\Gamma}_{\alpha}$ is the broadening matrix of lead $\alpha$, due to the coupling of the device to this lead. See Ref.~\cite{Sanz2023} for further details on the implementation.
From the transmission probability one can obtain the zero-bias conductance, calculated as
\begin{equation}\label{eq:conductance}
    G^{\sigma}_{\alpha\beta} = \text{G}_0\sum_{n}T^{\sigma n}_{\alpha\beta}(E_F),
\end{equation}
where $\text{G}_0$ is the conductance quantum and $T^{\sigma n}_{\alpha\beta}(E_F)$ is the transmission of the $n$th available channel at the Fermi level $E_F$, which is related to \Eqref{eq:transmission} by $T_{\alpha\beta}(E)=\sum_n T^{\sigma n}_{\alpha\beta}(E)$.
Note that around $E_F$ there is only one single transverse mode (channel) available, and therefore $T^{{\sigma}n}_{\alpha\beta}(E)=T^{\sigma}_{\alpha\beta}(E)$.
To compute the transmission probabilities we use the open-source code \textsc{TBtrans} \cite{Papior2017} and the python package \textsc{sisl} for post-processing \cite{zerothi_sisl}.

\section{Results}
\label{sec:results}
\begin{figure}
    \centering
    \includegraphics[width=\columnwidth]{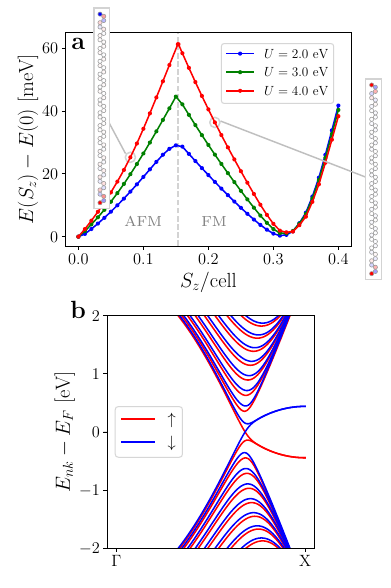}
    \caption{(a) Energy differences between MFH solutions calculated with $U=2$ eV (blue line), $U=3$ eV (green line) and $U=4$ eV (red line), obtained by imposing different spin projections $S_z$ per unit cell. The dashed line separates the two phases depending on $S_z$ (AFM and FM). The inset figures show examples of the spin polarization for the AFM and FM configurations, calculated with $S_z=0.08$ and $S_z=0.21$, respectively, where the red color indicates up-spin majority while the blue color indicates down-spin majority. (b) Band structure of the periodic 30-ZGNR calculated with $U=3.0$ eV for $S_z = 0.317$.
    Red and blue lines represent the up- and down-spin components, respectively. }
    \label{fig:30ZGNR-bands}
\end{figure}

\begin{figure}
    \centering
    \includegraphics[width=\columnwidth]{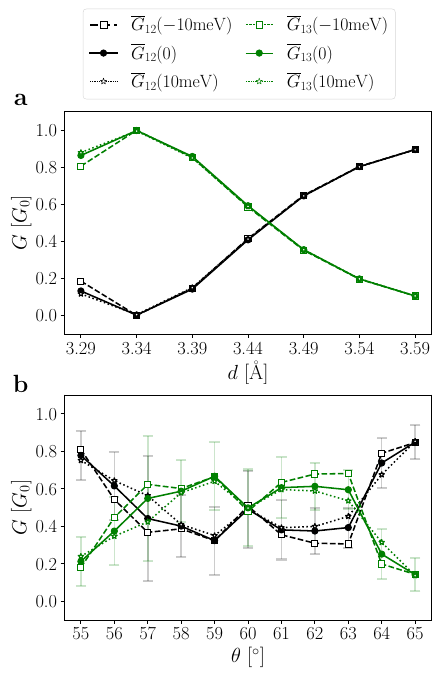}
    \caption{Spin-averaged conductance $\overline{G}_{\alpha\beta}(V)$ between incoming electrode $\alpha=1$ and outgoing electrodes $\beta=$2 (black lines) and $\beta=3$ (green lines) in units of the conductance quantum $G_0$, as a function of (a) the inter-layer separation $d$, with fixed crossing angle $\theta=60^{\circ}$ and stacking as shown in \Figref{fig:device}, and (b) the crossing angle $\theta$ averaged over the in-plane translations of one ribbon with respect to the other, with fixed $d=3.34$ {\AA}, for a device formed of crossed 30-ZGNRs obtained with $U=3.0$ eV in the FM configuration. The error bars in panel (b) are calculated as the standard deviation of $\overline{G}_{\alpha\beta}(0)$ at each $\theta$ by averaging over the different displacements. We obtain this conductance at different gate voltages $V=-10$ meV (dashed lines with open squares), $V=0$ (solid lines with filled circles) and $V=10$ meV (dotted lines with open stars). The legend placed on top is common to both panels (a,b).}
    \label{fig:30ZGNR-conductance}
\end{figure}
In this section we analyze the transport properties for a device formed of two crossed ZGNRs of $W=30$ carbon atoms across (30-ZGNR) as a function of the inter-layer separation $d$ for values close to the typical distance between layers in graphite ($d=3.34$ {\AA}), and the intersecting angle $\theta$ for values close to the commensurate case where $\theta=60^{\circ}$.
To understand the spin states of ZGNRs, we performed different spin-polarized calculations changing the total mean value of the spin operator $\hat{S}_z$ per unit cell, $\langle \hat S_z\rangle=\frac{1}{2}\sum_{i}\left(\langle n_{i\uparrow} \rangle - \langle n_{i\downarrow} \rangle\right) \equiv S_z$, where the summation goes over the sites $i$ within the unit cell of the periodic ZGNR.

In \Figref{fig:30ZGNR-bands}(a) we show the total energy per unit cell as a function of $S_z$ relative to the case of $S_z=0$ (the AFM case) for a periodic ZGNR of $W=30$ carbons across. As it can be seen here, there is a local minimum at $S_z = 0.317$, corresponding to the solution of lowest energy for $S_z\neq 0$.
The fact that the solution of minimum energy appears at such total $S_z$ can be understood from the fact that in the AFM case, the local spin projection summed over the bottom (or top) half of the unit cell of the ZGNR is $|S_z^\text{half}|=0.159$.
This means that the total $S_z$ per unit cell in the FM case needs to reach two times this value to flip the local magnetic moment at one edge.
Note that the magnetic moment associated to $S_z$ is $\mu=g_S\mu_{B}S_z$, where $g_S\approx 2$ is the electron spin $g$-factor and $\mu_B$ is the Bohr magneton.
To see to what extent the ribbon width affects these results, we compare $E(S_z)$ for $W=10, 20, 30, 40$-ZGNRs in Fig.~S1 in the supplemental material (SM). There we observe two main features: While the qualitative behavior is the same for all of them, the value of $S_z$ at which the minimum of energy appears is larger for wider ribbons, and, as expected, the minimum value of $E(S_z>0)$ diminishes with the width.

For each $S_z$ we plot the energy corresponding to the spin configuration of lowest energy in \Figref{fig:30ZGNR-bands}(a).
Here we distinguish between two phases depending on $S_z$: AFM character (for $S_z<0.15$), where the spin polarization shows opposite spin majorities at the edges, and FM character (for $S_z>0.15$), where the spin polarization shows spin majority of equal spin index. 
The two insets to \Figref{fig:30ZGNR-bands}(a) show the spin polarization for a 30-ZGNR: One in the AFM-like spin configuration (calculated with $S_z=0.08$), where it can be seen that the colors at the edges are different (red and blue), and another one in the FM-like spin configuration (calculated with $S_z=0.21$), where it can be seen that the same color appears at both edges (red).
In the case of the AFM-like spin configuration for $S_z\neq 0$, not only the sign of the local magnetic moments at the bottom and top edges of the unit cell is different but also the magnitude, as a consequence of the existing spin imbalance. Whereas when the FM character is achieved, both the magnitude and sign of the local magnetic moments at the bottom and top edges of the unit cell are equal.

In \Figref{fig:30ZGNR-bands}(b) we plot the band structure for the FM solution of lowest energy for the 30-ZGNR, obtained with $S_z = 0.317$, for spin $\sigma=\uparrow$ (red lines) and $\sigma=\downarrow$ (blue lines). Here we can observe the metallic character of the FM configuration for the ZGNR, as there are states available at the Fermi level, $E_F$, for both up and down spins.

As mentioned above, although the ground state corresponds to the configuration with $S_z=0$, the presence of a magnetic field $B$ in $z$-direction (cf.~\Figref{fig:device}) can stabilize a high-spin configuration due to the Zeeman energy $\Delta E = \mu B= g_S\mu_{B}S_z B$.
For instance, the corresponding electronic energy $E(S_z)$ for the FM-like configuration of lowest energy is $E(S_z=0.317)=0.97$ meV/cell above the ground state, implying that a critical magnetic field of the order $B_c=26.6$ T (parallel to the $z$-axis in this case) is needed to make the two spin states degenerate.%
In \Figref{fig:30ZGNR-conductance}a we study the zero-bias conductance $G_{\alpha\beta}(V)$ with $(\alpha,\beta)\in \lbrace (1,2), (1,3) \rbrace$ (black and green lines, respectively) for a device formed of two crossed 30-ZGNRs as a function of the inter-layer separation $d$. Here, $V$ represents a rigid shift of the Fermi level $E_F$. We consider inter-layer distances close to the typical van-der-Waals distance between graphene layers in graphite ($d=3.34$ {\AA}) \cite{Brandimarte2017, PhysRev.100.544, PhysRevB.40.993}.
In the first place, we can infer that the total spin-averaged conductance (sum of intra- and inter-layer conductances) is 1 since the values for $\overline{G}_{12}$ and $\overline{G}_{13}$ are symmetric with respect to $0.5\text{G}_0$, which means that there is no backscattering for an incoming electron at the Fermi level in these devices at least for these ranges of $d$ and $\theta$.
In the second place, we observe an oscillating behavior of $\overline{G}_{\alpha\beta}$ with respect to this varying parameter.
For instance, the inter-/intra-layer conductance ratio reaches its maximum for $d=3.34$ {\AA}. While one would expect that for smaller inter-layer distances $d$ the interlayer ($\overline{G}_{13}$) conductance would increase, as the interlayer hopping integral depends exponentially on the distance between the ribbons, we observe a decrease (and increase of $\overline{G}_{12}$) for smaller $d$ in \Figref{fig:30ZGNR-conductance}(a), as a consequence of an interference process due to the scattering potential created by the crossing.
We also observe that for $d$ between 3.44 and 3.49 {\AA} there is a crossing between $\overline{G}_{12}$ and $\overline{G}_{13}$, implying that for that inter-layer separation the device behaves as a perfect 50:50 beam splitter where the incoming electron beam is equally separated in the two possible outgoing directions with $\overline{G}_{\alpha\beta}=0.5\text{G}_0$ for low gate voltages $V$.

Similarly, in \Figref{fig:30ZGNR-conductance}b we study  $G_{\alpha\beta}(V)$ for different crossing angles close to the commensurate configuration with $\theta=60^{\circ}$.
We apply the rotation around the center of the
scattering region (crossing) that is obtained for the case with $\theta=60^{\circ}$, and account for the effect of different possible stackings by averaging over the in-plane translations of one ribbon with respect to the other. By doing so, we aim to provide a comprehensive overview of the results, accounting for the variability in stacking configurations that might occur in practical scenarios. The in-plane unit cell is determined by the graphene lattice vectors. We obtain the conductance for a mesh of four points along each lattice vector within the unit cell. The error bars are calculated as the standard deviation of the spin-averaged conductance $\overline{G}_{\alpha\beta}$ at each point, averaging over the in-plane translations. The observed variance of approximately $\sim$10-20\%  reflects the variations across different translational configurations, showing the inherent differences sampled by these translations. However, not all the stackings are equivalent. For instance, the most energetically favourable (and therefore most likely) configuration is the AB-stacking (see the SM of ref. \cite{Sanz2022}).
Analyzing the transport properties relative to this varying parameter in \Figref{fig:30ZGNR-conductance}b we observe, on one hand, that the inter-/intra-layer conductance ratio reaches its maximum for $\theta=55^{\circ}, 65^{\circ}$. On second hand, the sum of the total spin-averaged conductance is 1 as in panel (a), since the values for $\overline{G}_{12}$ and $\overline{G}_{13}$ are symmetric with respect to $0.5\text{G}_0$ as well, meaning that the variation of $\theta$ does not introduce backscattering.
We can see that the oscillatory dependence of the conductance on the crossing angle is less smooth than the one seen in \Figref{fig:30ZGNR-conductance}a. This occurs due to a more complicated dependence of the $\sigma$-and $\pi$-type hopping integrals on $\theta$.

To see what is the effect of the width on the transport properties as a function of these two varying parameters we performed a similar analysis for a 20-ZGNR device in the SM (see Fig.~S2), where we observe that, qualitatively, the behavior is maintained.
For further detail we plot the energy-resolved transmission probabilities for the 30-ZGNR device as a function of $d$ and $\theta$ in the SM (see Figs.~S3 and S4).

Finally, we note that it has been previously shown that the symmetries associated to the spatial distribution of the spin densities are crucial for the transport properties of the device \cite{Sanz2020, Sanz2022}. In this case, since the FM character implies that $\langle n_{\uparrow}\rangle \neq \langle n_{\downarrow}\rangle$, there will not be a symmetric behavior for the existing spin channels. However, the spin density distribution possesses a symmetry axis at $y=\sin(-60^{\circ})x$ that maps the device geometry to itself through mirror operations, and applies to each spin component individually (conserves the spin index). As it has been shown in references \cite{Sanz2020, Sanz2022}, there are certain symmetrical combinations of electrodes which lead to equal transmission probabilities $T^{\sigma}_{\alpha\beta}=T^{\sigma}_{\gamma\delta}$. In this case, the symmetrical electrodes mapping corresponds to $(1,2,3,4) \leftrightarrow (4,3,2,1)$.

\section{Conclusions}
\label{sec:conclusions}

We have analyzed the electron transport properties for a device formed of two crossed infinite ZGNRs of $W=30$ carbon atoms across (30-ZGNRs) as a function of the spin configuration by fixing different values for the total spin per unit cell $S_z$. In first place, by computing the total energy associated to these configurations $E(S_z)$ we have shown that there is a local minimum for the solution with $S_z>0$, with $E(S_z>0)$ close to 1 meV/cell above the ground state ($E(0)$). We have also seen that depending on $S_z$ there are two possible phases: AFM-character, where the edges of the ZGNR unit cell are populated by opposite spin majorities, and FM-character, where the two edges of the ribbon are populated by the same spin majority. These two phases appear for $S_z<0.15$ and $S_z>0.15$, respectively. We also computed the band structure for the FM-like configuration of lowest energy, where we observe that this system in such spin state shows metallic character.
We estimate that the critical magnetic field needed to make this FM-like solution degenerate with the AFM ground state is $B_c=26.6$ T for this particular case, although this value will further decrease for wider ribbons.

We have also calculated the inter- and intra-layer electrical conductances for different gatings varying the inter-layer distances, for distances close to the van-der-Waals distance between graphene layers in graphite ($d=3.34$ {\AA}), and crossing angles close to the commensurate stacking where $\theta=60^{\circ}$ for this 4-terminal device. 
We have shown that the (spin- and displacement-averaged) electrical conductance displays an oscillatory behavior with respect to these varying parameters at low gate voltages ($-10 \mathrm{meV}\leq V \leq 10 \mathrm{meV}$) while maintaining the sum $\overline{G}_{12}+\overline{G}_{13}=1$, which means that there is no backscattering for the devices for different values of $d$ and $\theta$ within the shown ranges nor conductance into terminal 4. The maximum value for the inter-/intra-layer spin-averaged conductance ratio ($\overline{G}_{13}/\overline{G}_{12}$) for this device is found for $d=3.34$ {\AA} and $\theta=55^{\circ}, 65^{\circ}$.
Additionally, to show that these results are not exclusive for the chosen ZGNR width, we performed a similar analysis for a 20-ZGNRs device (see SM), where we show that it possesses similar qualitative behavior.

The results presented here add to the vision of using GNR-based devices for spintronics and quantum technologies. On top of the already discussed properties and applications of spin-polarized GNR-based beam splitters for electron quantum optics \cite{Sanz2020, Sanz2022, Sanz2023}, this device in its FM-like spin configuration can be a promising candidate given its metallic character, since electron injection can be considerably facilitated as one can generate a minimal excitation in the device by applying a small voltage pulse \cite{Levitov1996, Ivanov1997, Keeling2006, Dubois2013}.

\section*{Acknowledgements}
This work was funded by the Spanish MCIN/AEI/ 10.13039/501100011033 (PID2020-115406GB-I00,  TED2021-132388B-C44, PID2022-140845OB-C66, and JDC2022-048665-I), the Basque Department of Education (PIBA-2023-1-0021), the University of the Basque Country (UPV/EHU) through Grant IT-1569-22, and the European Union’s Horizon 2020 (FET-Open project SPRING Grant No.~863098).

%



\section*{Supplementary material}

## S2. TRANSMISSION PROBABILITIES

In Fig. S3 we plot the intra- and inter-transmission probabilities for an electron incoming the device from terminal 1 (see Fig. 1 of main text) for different inter-layer distances (a) $d = 3.34$, (b) $d = 3.39$, (c) $d = 3.44$ and (d) $d = 3.49$ Å. In Fig. S4 we perform a similar analysis for different crossing angles (a) $\theta = 56°$, (b) $\theta = 58°$, (c) $\theta = 60°$ and (d) $\theta = 62°$. Transmission into terminal 4 and reflection (losses) are not shown here since they remain zero for the single-mode energy region [1, 2], as it can be inferred from the fact that $T_{12}^\sigma + T_{13}^\sigma = 1$ in this energy window. By comparing panels (a-d) in Fig. S3 it can be seen that the inter-/intra-transmission ratio close to $E_F$ ($T_{13}^\sigma/T_{12}^\sigma$) diminishes with the distance. While for $d = 3.34$ Å this ratio lies close to 1/0 for $-0.1 < E < 0.2$ ($-0.2 < E < 0.1$) eV for $\sigma = \uparrow$ ($\sigma = \downarrow$), when the inter-layer distance is $d = 3.49$ Å, the device splits the beam almost equally into arms 2 and 3 close to the Fermi level.

In Fig. S4 we can observe that the crossing angle also has an effect on the $T_{13}^\sigma/T_{12}^\sigma$ ratio, as expected [1]. Here we see that for angles $\theta \neq 60°$ there is a reduction of the $T_{13}^\sigma/T_{12}^\sigma$ ratio. On the other hand, it is worth mentioning that, while in the commensurate case $\theta = 60°$ we find an approximate symmetry between $T_{\alpha\beta}^\sigma(E)$ and $T_{\alpha\beta}^{\bar\sigma}(-E)$ (as seen in Fig. S3), for different angles this quasi-symmetrical behavior is lost. A similar behavior was found for unpolarized devices where the approximate particle-hole symmetry found for the commensurate pattern disappeared for non-commensurate stackings [1].

---

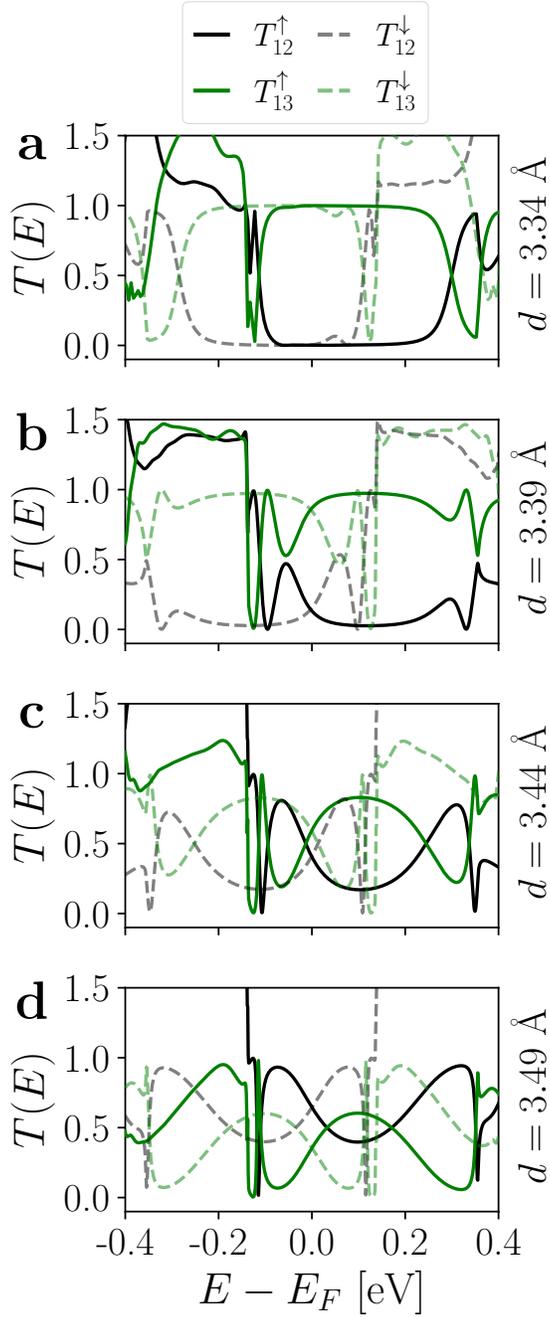

FIG. S3. Transmission probabilities between incoming electrode 1 and outgoing electrodes 2 (black lines) and 3 (green lines), as a function of electron energy for crossed 30-ZGNRs obtained with $U = 3.0$ eV in the FM configuration for a device with inter-layer separations (a) $d = 3.34$, (b) $d = 3.39$, (c) $d = 3.44$ and (d) $d = 3.49$ Å for electrons with spin index $\sigma =\uparrow$ (solid lines), $\sigma =\downarrow$ (dashed lines). The legend placed on top is common for all panels (a-d).

5

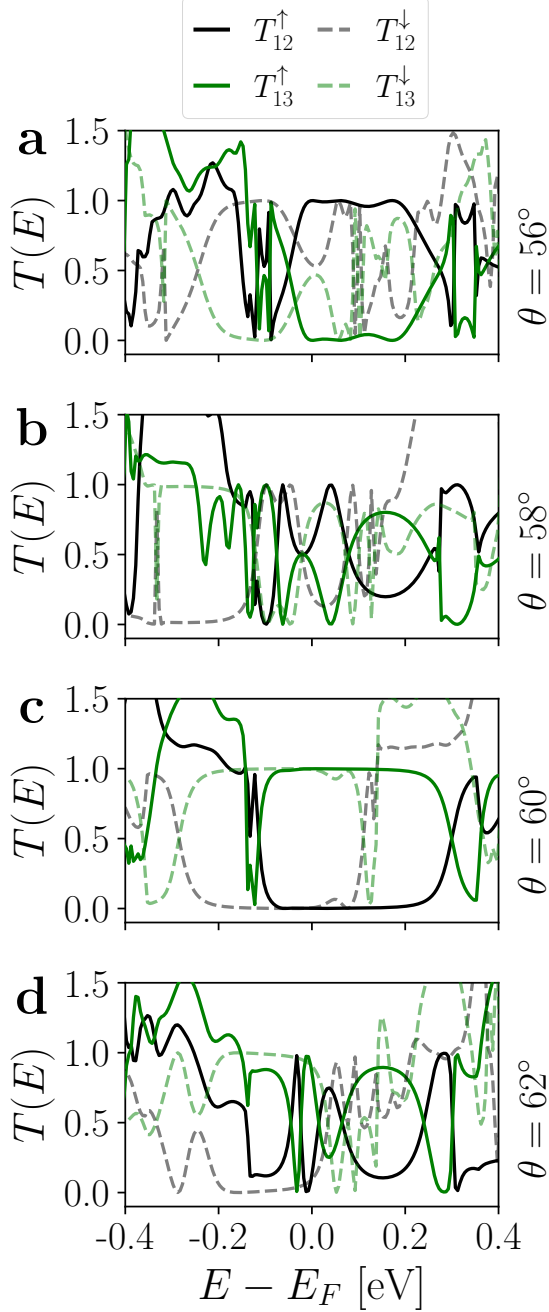

FIG. S4. Transmission probabilities between incoming electrode 1 and outgoing electrodes 2 (black lines) and 3 (green lines), as a function of electron energy for crossed 30-ZGNRs obtained with $U = 3.0$ eV in the FM configuration for a device with fixed inter-layer distance $d = 3.34$ Å and varying intersecting angle (a) $\theta = 56°$, (b) $\theta = 58°$, (c) $\theta = 60°$ and (d) $\theta = 62°$ for electrons with spin index $\sigma =\uparrow$ (solid lines), $\sigma =\downarrow$ (dashed lines). The legend placed on top is common for all panels (a-d).

6


\thispagestyle{empty}

\end{document}